\documentclass{PoS}

\title{Present status of light flavoured scalar resonances}

\ShortTitle{Present status of light flavoured scalar resonances}

\author{\speaker{J. R. Pel\'aez}\\
        Departamento de F\'{\i}sica Te\'orica II. Facultad de CC. F\'{\i}sicas. Universidad Complutense. 28040 Madrid. SPAIN\\
        E-mail: \email{jrpelaez@fis.ucm.es}}

\abstract{ This talk is a brief account of recent developments on light scalar meson spectroscopy. I mostly focus on the major revision of the $\sigma$ or $f_0(500)$ meson in the Review of Particle Physics, driven by new data and the accummulation of rigorous dispersive studies, but I also comment on recent updates of other scalars and future progress. }

\FullConference{The 7th International Workshop on Chiral Dynamics,\\
		August 6 -10, 2012\\
		Jefferson Lab, Newport News, Virginia, USA}

\begin{document}

\section{Introduction}
\vspace*{-.3cm}

Despite playing
a prominent role in the attractive part of the nucleon-nucleon interaction, the QCD spontaneous chiral symmetry breaking and the search for glueballs, light scalars have suffered a longstanding debate
about their properties, spectroscopic classification and even their very existence---see the ``Note on light scalars below 2 GeV'' in the Review of Particle Properties (RPP) \cite{PDG}. However, the combination
of rigorous and model independent approaches with new data has provided very convincing proof of the existence and properties of these states. Slowly and cautiously, these developments are being reflected in the RPP, which in its latest edition has finally made a major revision 
on the $\sigma$ meson and updated other resonances. 

Here I will briefly comment on progress, made after the previous 2009 Chiral Dynamics Workshop, on mass and width determinations of 
scalars below 1 GeV.
 Other scalars and the heated controversy concerning their classification in multiplets and composition, lie beyond the scope of this mini-review.
I will follow two paths: a conservative one, based on the RPP updates, and my personal view, less conservative but probably closer to the ``scalar community'', for long well  aware of the situation now acknowledged by the RPP revisions. I will explain how the RPP updates have been driven, not only by new data, but by the consistency of rigorous dispersive approaches. Since such analyses exist for other light scalars, I expect further revisions in the near future.

\vspace*{-.3cm}
\section{The $\sigma$ or $f_0(500)$ meson. A major change in the RPP. }
\vspace*{-.2cm}

A light scalar-isoscalar field was postulated 60 years ago \cite{Johnson:1955zz}, to explain the nucleon attraction, and was soon incorporated in the Linear Sigma Model \cite{GellMann:1960np}, from which it gets its common name: the $\sigma$ resonance. Nowadays it is called $f_0(500)$. 
Being linear, this is the simplest realization of an spontaneous chiral symmetry breaking,
through a scalar multiplet, where all fields but the $\sigma$ become Goldstone bosons.
On more general grounds, the $\sigma$, with the vacuum quantum numbers, 
is expected to play a relevant role in the QCD spontaneous chiral symmetry breaking. 

\begin{figure}[t]
\vspace*{-2.1cm}
\hspace*{-.6cm}
  \includegraphics[width=.54\textwidth]{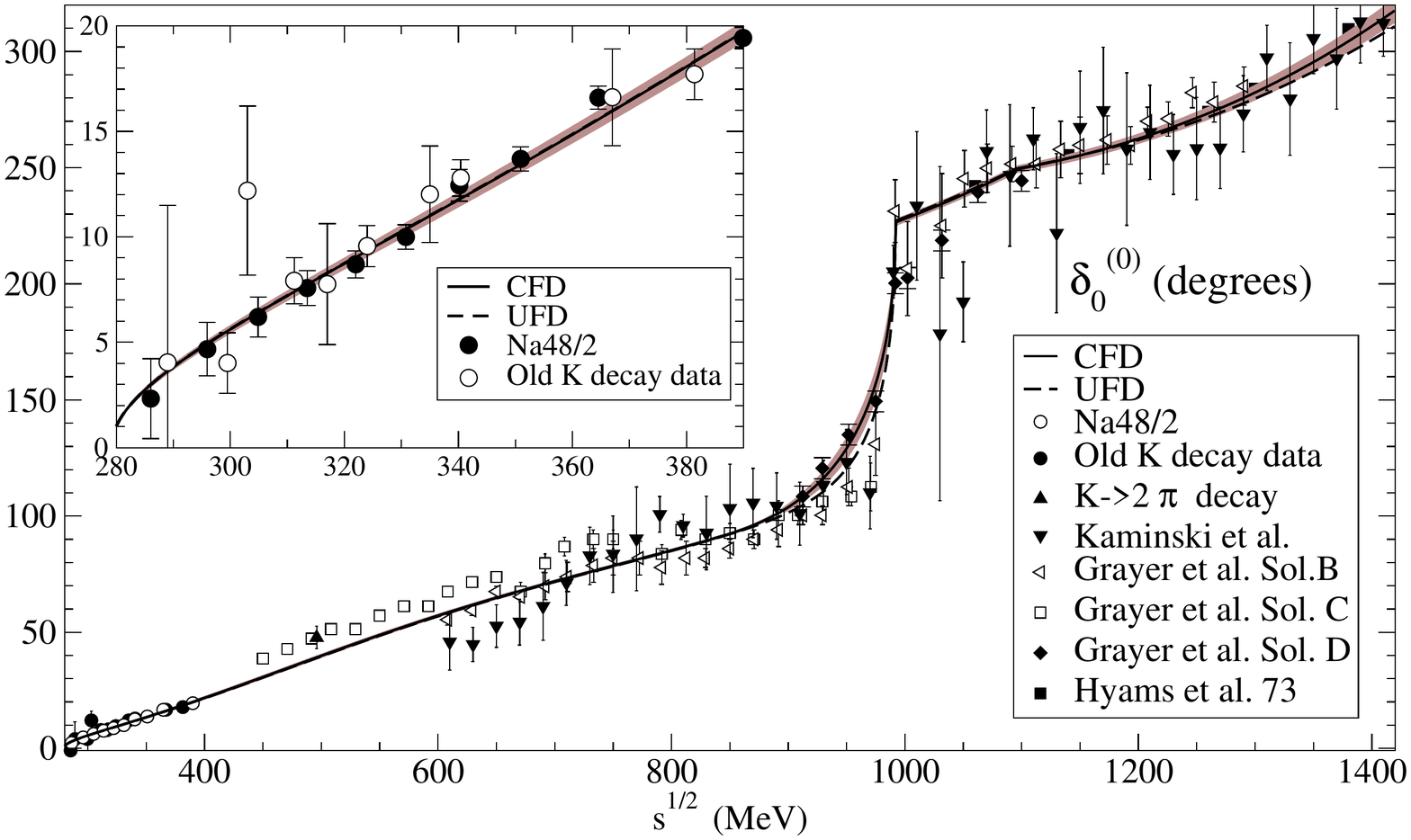}
\hspace*{-.7cm}
  \includegraphics[width=.53\textwidth]{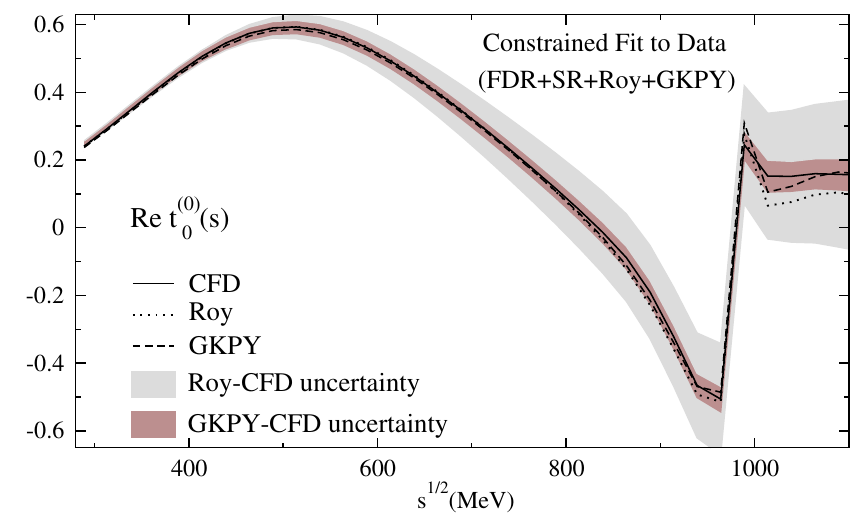}
\vspace*{-1cm}
  \caption{ Scalar-isoscalar $\pi\pi$ scattering. Left: Data on the $\delta_0^{(0)}$ scattering phase \cite{Pr73,Rosselet:1976pu,Batley:2010zza} versus the UFD and CFD parametrization \cite{GKPY11}.
Right: Fulfillment of Roy and GKPY equations for this same wave. 
}
  \label{fig:UFDCFD}
\vspace*{-.5cm}
\end{figure}

The significance of the latest RPP major revision of the $\sigma$ meson
can be illustrated as follows: until 1974 the $\sigma$ was listed as ``not-well established'', disappeared for 20 years after 1976 and came back as the $f_0(600)$ in 1996. 
The cause of this coming in, out and back to the tables is that nucleon-nucleon interactions are not sensitive to the details of the exchanged
particles, even less so if they are as wide as the $\sigma$ and thus light scalars were studied in meson-meson scattering, where they can be produced in the s-channel. Unfortunately, $\pi\pi$ scattering is extracted from $\pi N\rightarrow \pi\pi N$ through a complicated analysis plagued with systematic uncertainties, and experiments \cite{Pr73} produced conflicting data sets. For instance, note in Fig.\ref{fig:UFDCFD}, which shows the scalar-isoscalar $\pi\pi$ scattering phase, the large differences between data sets \cite{Pr73}, even within the same collaboration.  Strong support for a $\sigma$ below 1 GeV came from heavy meson decays, making the $f_0(600)$ case sufficiently convincing to be considered ``well established'' in 2002, although with a huge mass uncertainty ranging from 400 to 1200 MeV and a similarly large range, from 500 to 1000 MeV, for the width. These huge ranges and the $f_0(600)$ name were kept until the last 2012 RPP edition.

\begin{figure}[t]
\vspace*{-.8cm}
  \centering
  \includegraphics[width=.49\textwidth]{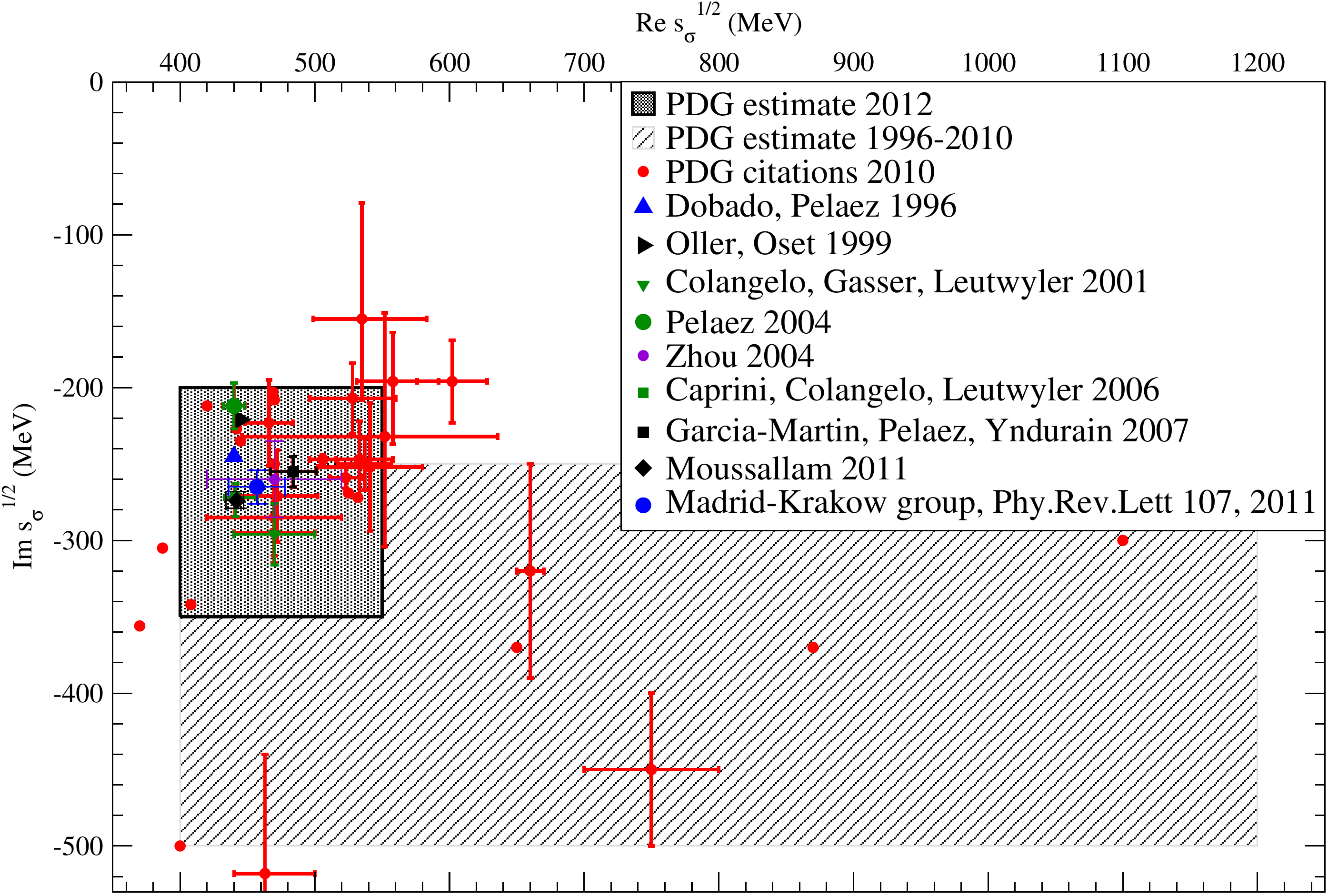}
  \includegraphics[width=.49\textwidth]{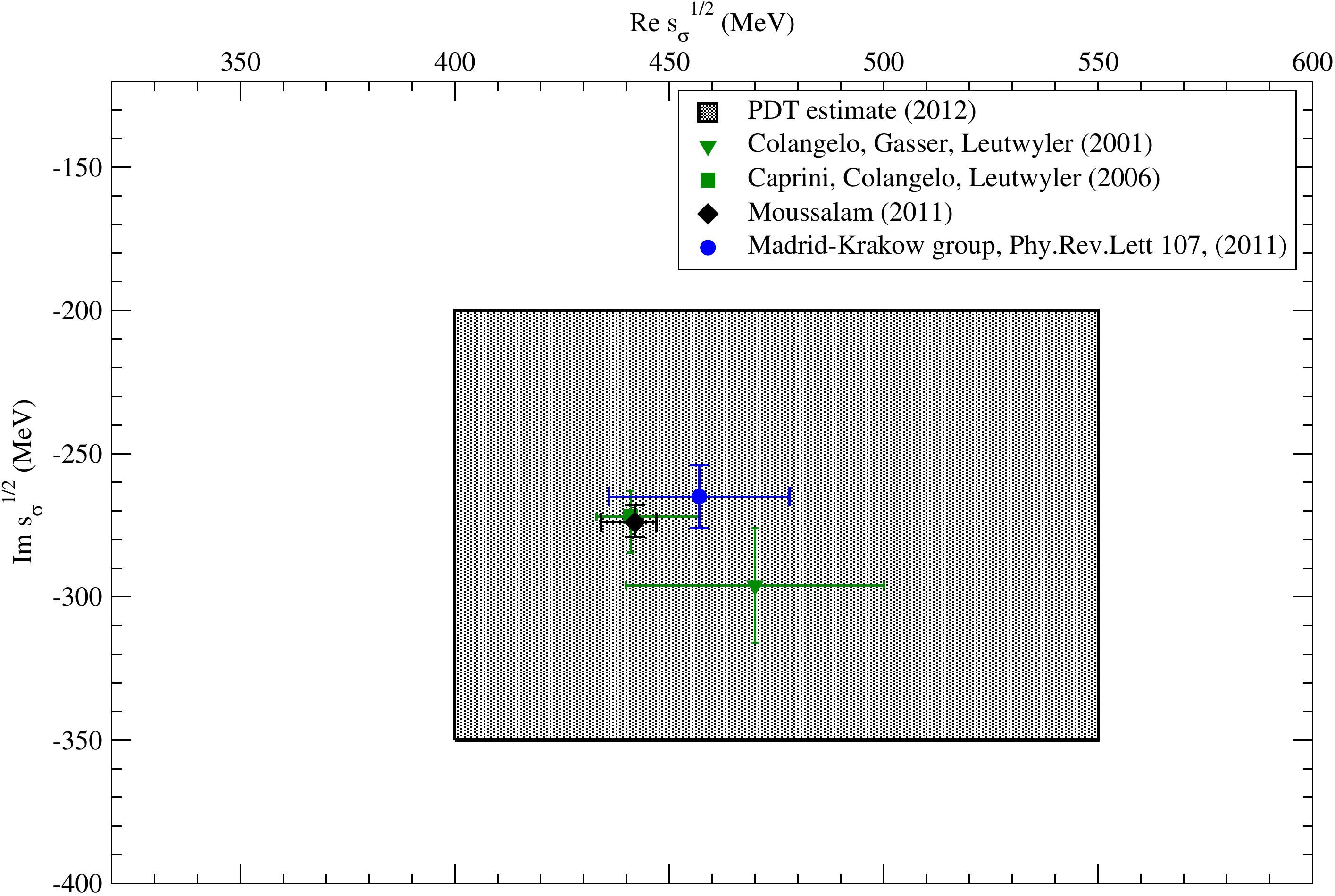}
  \caption{Left: $f_0(500)$ poles in the RPP\cite{PDG}. 
Non-red poles, obtained from dispersive or analytic approaches \cite{others,CGL,Caprini:2005zr,GarciaMartin:2011jx,Moussallam:2011zg}, fall within the 
2012 estimate, which is a major revision
with respect to the 2010 RPP estimate.
Right: The four ``most advanced dispersive analyses'' \cite{CGL,Caprini:2005zr,GarciaMartin:2011jx,Moussallam:2011zg} which according to the ``Note on light scalars'' of the 2012 RPP lead to
 their ``more radical'' average $\sqrt{s_\sigma}=(446\pm6)-(276\pm5)\,$ MeV.  
}
  \label{fig:poles}
\end{figure}

Two remarks are in order about Fig.\ref{fig:UFDCFD}. 
First, note the data below 400 MeV coming from $K\rightarrow\pi\pi\ell\nu$ decays \cite{Rosselet:1976pu,Batley:2010zza}, which have almost no systematic uncertainty compared to those from $\pi N\rightarrow\pi\pi N$. Especially relevant are the 2010 precise NA48/2 data \cite{Batley:2010zza}, since consistency with them is a key requirement for the RPP choice of results in their new estimate. 
Second, no Breit-Wigner shape is seen around 500-600 MeV. Actually, the $\sigma$  cannot be described as a Breit-Wigner resonance (nor the $K_0^*(800)$).  Hence one uses
the mathematically rigorous definition of a resonance by means of its associated pole in the complex plane, whose position $s_R$ is related to the resonance mass and width as $\sqrt{s_R}\simeq M_R-i \Gamma_R/2$. This is why the RPP provides the so-called ``t-matrix'' pole, although, unfortunately, it also provides a Breit-Wigner pole. To my view, the latter only leads to confusion, and I will only comment ``t-matrix'' poles. Thus, Fig.\ref{fig:poles} shows the position of the $\sigma$ poles in the RPP and the huge light gray area corresponds to the RPP estimate until 2010.

Let me emphasize that to determine poles deep in the complex plane, 
a good data description is not enough, but also
a correct analytic continuation is needed. Unfortunately this is not the case of many analyses, leading to poor or plain wrong pole determinations. Indeed, a significant part of the disagreement seen in Fig.\ref{fig:poles} is due to unreliable extrapolations to the complex plane, so that even the same experiment can provide dramatically different poles. For example, the poles at 400-i 500 MeV, 1100-i300 MeV and 1100-i 137 MeV (below the legend), all come from 
\cite{Amsler:1995bf}. To my view, only poles extracted from analytic or 
dispersive approaches provide reliable $\sigma$ pole determinations, which are highlighted in colors other than red in Fig.\ref{fig:poles}. It is clear that, within the light scalar community, the existence of a $\sigma$ pole 
around 500 MeV was rather well known for many years. 
Dispersive analyses may differ by few tens of MeV, not several hundreds.
 Poles determined from heavy meson decays ( no updates in the 2012 RPP \cite{PDG}) yield a somewhat higher mass than dispersive approaches, between 500 and 550 MeV. Unfortunately
the analysis of these decays has been usually performed with models less rigorous than dispersive approaches. 

 A rigorous analytic continuation is obtained from dispersion relations that, as a consequence of causality, relate the amplitude at any value with an integral over the real axis, i.e. the data.Thanks to the integral representation the results are independent of the model or
functional form parametrizing the data. They can be used to: a) check the consistency of data at a given energy against data in other regions, b) constrain data fits, c) calculate the amplitude at energies where data do not exist, c) use the analytic continuation to look for poles. 
Of particular interest for spectroscopy are partial wave dispersion relations, 
since their poles are directly associated to resonances with their same quantum numbers. 
However, due to crossing symmetry, partial waves have a ``left cut''  contribution, 
from the unphysical $s$ region. This is numerically relevant for precise studies of 
the $\sigma$ and the $K_0^*(800)$, which are relatively close to threshold and the left cut. Dealing rigorously with the left cut involves an infinite set of coupled integral equations, known as Roy equations \cite{Roy:1971tc} for $\pi\pi$ scattering, which have received considerable attention over the last decade \cite{ACGL,CGL,Kaminski:2002pe,Caprini:2005zr,GKPY11,GarciaMartin:2011jx,Moussallam:2011zg}.  In the 70's, their accuracy was limited by the quality of threshold data, but this caveat can be circumvented either by the use of Chiral Perturbation Theory (ChPT) at low energy, as in \cite{CGL},
or, if one wants to avoid the use of ChPT as in \cite{GKPY11}, by using the recent and precise NA48/2 data \cite{Batley:2010zza}. The former approach provided a precise $\sigma$ pole, also showing that Roy Equations yield a consistent analytic continuation to the $\sigma$ region \cite{Caprini:2005zr}. The latter, which is just a dispersive data analysis, was followed by our group \cite{GKPY11,GarciaMartin:2011jx} using another set of Roy-like Equations, called GKPY Equations,  with only one subtraction (less energy suppression in the integrals).

Thus, the 2012 RPP has finally made a major revision of the $\sigma$ 
mass range, reduced by a factor of more than five, down to 400 to 550 MeV,
 and by almost a factor of two for the width, now estimated between 400 and 700 MeV.
This change has been triggered by the consistency of dispersive results, together with the new  NA48/2 data close to $\pi\pi$ threshold \cite{Batley:2010zza}.
The new RPP ``estimate'', shown in Fig.\ref{fig:poles} as the smaller and darker rectangle, takes into account, not only the most recent dispersive analyses, but 
other results from models which are required to be at least consistent with the accurate $K\rightarrow\pi\pi\ell\nu$ data \cite{Batley:2010zza,Pislak:2001bf}, as well as values from other processes like heavy meson decays, which, as commented above, yield somewhat larger masses than dispersive approaches, extracted using models. 
Note that even the name of the particle has been changed to $f_0(500)$.
The RPP also provides Breit-Wigner parameters but I have argued above why I think they should be  avoided.
Admittedly, this major revision constitutes a very considerable and long awaited improvement upon the previous situation. But, to my view, these RPP criteria are still rather conservative, and for the $\sigma$ I would only rely on pole extractions based on rigorous analytic methods.  Actually, even the RPP `Note on light scalars'' suggests that one could 
``take the more radical point of view and just
average the most advanced dispersive analyses'' (which here correspond to \cite{CGL,Caprini:2005zr,GarciaMartin:2011jx,Moussallam:2011zg}, shown in Fig.\ref{fig:poles}), to find: $\sqrt{s_\sigma}=(446\pm6)-(276\pm5)\,$ MeV. 

Let me illustrate the dispersive techniques, by sketching the methods of our group \cite{GKPY11,GarciaMartin:2011jx}. We start from a set of ``Unconstrained Fits to Data'' (UFD), shown to be not too inconsistent with forward dispersion relations. Next, we slightly modify the fit parameters to satisfy dispersion relations without spoiling the data description, obtaining ``Constrained Fits to Data'' (CFD). Both the CFD and UFD for the scalar-isoscalar $\pi\pi$ scattering phase shift are shown on Fig.\ref{fig:UFDCFD}. The only sizable differences between the UFD and CFD appear in the 1 GeV region and above, but both describe the data very well. 
As an example, we show in Fig.\ref{fig:UFDCFD} how well the CFD satisfies the Roy and GKPY equations for the real part of the scalar-isoscalar wave. The continuous line is the CFD input, whereas the dotted and dashed lines are the output of the Roy and GKPY equations, respectively. Note that the once-subtracted GKPY equations are more precise in the resonance region, say above 500 MeV, whereas Roy equations are more accurate below that energy, given the same input.
With this CFD, which describes the data and is consistent with a whole set of dispersion relations, unitarity and symmetry constraints, etc... we use the dispersion relation to continue the amplitude into the complex plane, finding the following poles \cite{GarciaMartin:2011jx}:
$\sqrt{s_\sigma}=(457^{+14}_{-13})-i(279^{+11}_{-7})$ MeV (from \, GKPY \,eqs.) 
and $(445\pm25)-i(278^{+22}_{-18})$ MeV (from \,Roy \,eqs.).

Our results just above are two of the five new entries in the 2012 RPP edition. The other new entries are two results from an ``analytic K-matrix model'' in \cite{Mennessier:2010xg}:  $(452\pm13)-i(259\pm16)\, {\rm MeV}$ and $(448\pm43)-i(266\pm43)\, {\rm MeV}$, depending on what data sets and different variants of the K-matrix model are averaged. Finally, the other new result in the 2012 RPP is $(442^{+5}_{-8})-i(274^{+6}_{-5})\, {\rm MeV}$ from \cite{Moussallam:2011zg}. The latter is also based on Roy equations using as input for other waves and higher energies
the Roy equations output of \cite{CGL} and is therefore very consistent with the older result in \cite{Caprini:2005zr}: $\sqrt{s_\sigma}=(441^{+16}_{-8})-i(272^{+9}_{-12.5})\, {\rm MeV}$, which used ChPT input, as well as with that even older in \cite{ACGL}: $(452\pm13)-i(259\pm16)\, {\rm MeV}$. These last three results, based on Roy equations, together with our two results in the paragraph above, are precisely the ones considered by the RPP as the ``most advances dispersive analyses'', shown in Fig.\ref{fig:poles} here.

\vspace*{-.3cm}
\section{The $f_0(980)$ , $a_0(980)$ and $K_0^*(800)$}
\vspace*{-.2cm}

The $f_0(980)$ and $a_0(980)$ existence and their parameters have been much less controversial over time, since they are narrow and clearly seen in many processes. For example, in Fig.\ref{fig:UFDCFD} a Breit-Wigner-like shape over a background phase of about 100 degrees may be seen around 980 MeV, corresponding to the $f_0(980)$, although it should be distorted by the nearby $\bar KK$ threshold.

\begin{figure}[t]
\vspace*{-.7cm}
  \centering
  \includegraphics[width=.43\textwidth]{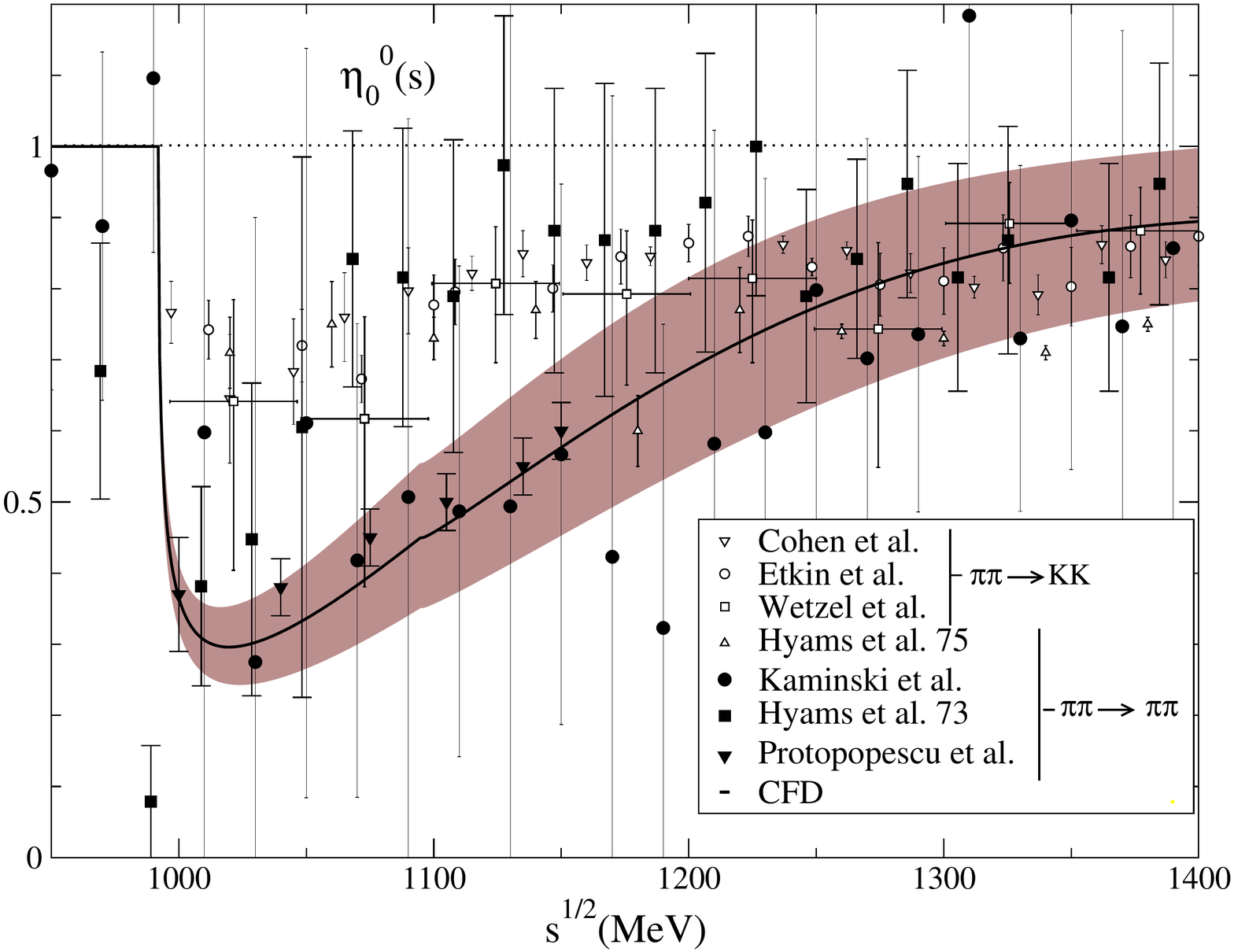}
  \includegraphics[width=.43\textwidth]{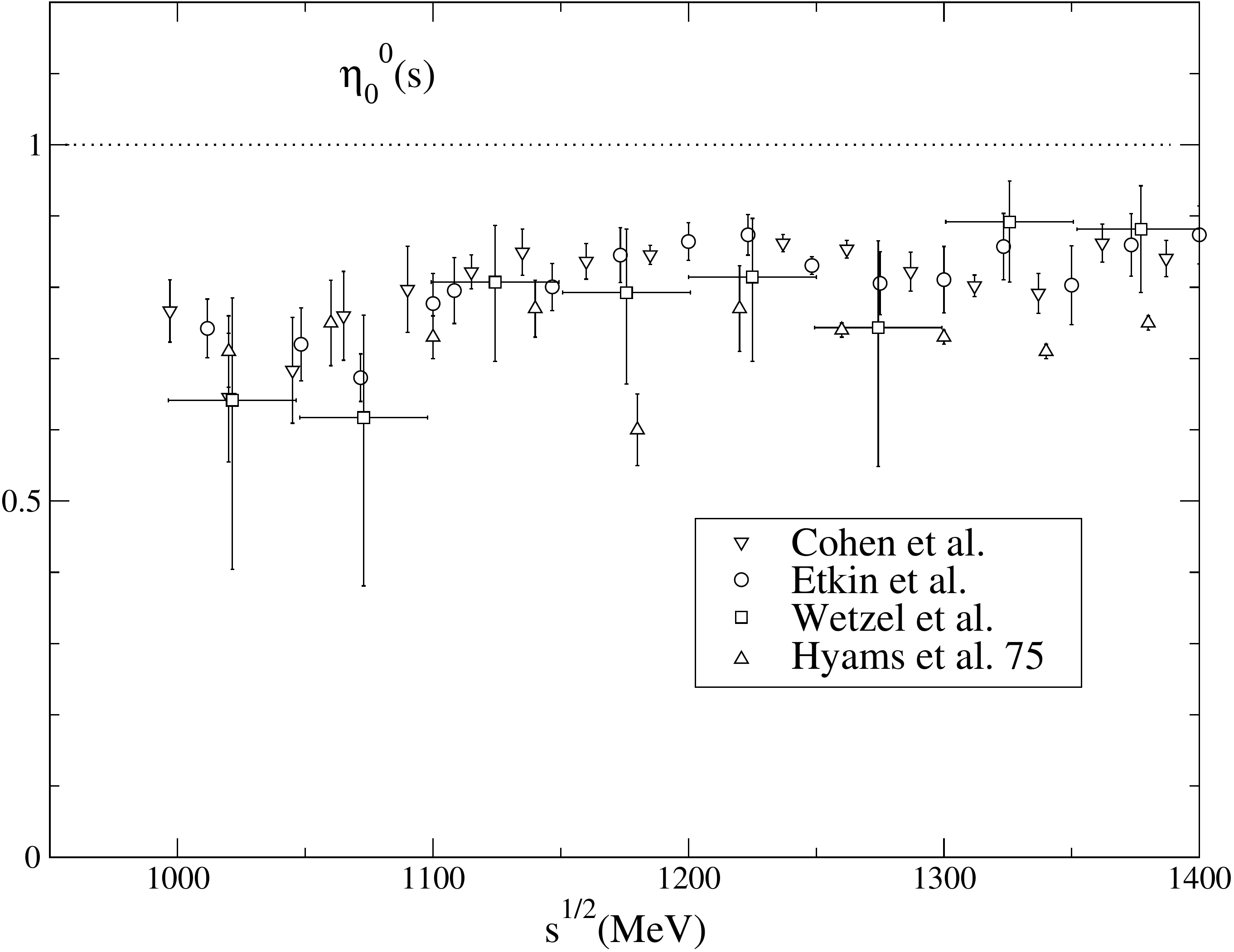}
\vspace*{-.3cm}
  \caption{ Left: Favored ``Dip'' solution (CFD) for the $\pi\pi$ scattering inelasticity around the $f_0(980)$ resonance region. Original figures and references in \cite{GKPY11}. Right: data only for the disfavored ``no-dip'' scenario. 
}
  \label{fig:S0inel}
\end{figure}

There have been no changes for the $a_0(980)$ in the latest RPP.
In contrast, after almost two decades of keeping the same estimate, the 2012 RPP has 
updated the $f_0(980)$ mass to $990\pm20$ MeV. As pointed out in the 2012 ``Note on light scalars'', 
the 10 MeV higher update on the mass, and the doubling of the uncertainty, was made to accommodate the dispersive analysis 
by our group \cite{GKPY11,GarciaMartin:2011jx}.
We obtain an $f_0(980)$ pole (in the second Riemann sheet) at: $\sqrt{s_{f_0(980)}}=(996\pm7)-i(25^{+10}_{-6})$ MeV if we use GKPY equations and $\sqrt{s_{f_0(980)}}=(1003^{+5}_{-27})-i(21^{+10}_{-8})$ MeV
from Roy equations. The relevance of our study is that it has settled a longstanding conflict between the ``dip'' and ``no-dip'' scenarios for the elasticity parameter in $\pi\pi$ scattering, shown in Fig.\ref{fig:S0inel}. In \cite{GKPY11} it was shown that the dip scenario satisfies well the GKPY dispersion relations as seen in  Fig.\ref{fig:S0inel}, whereas it is not possible to accommodate the non-dip scenario. This was confirmed later in \cite{Moussallam:2011zg} using Roy equations and obtaining a pole at: $(996^{+4}_{-14})-i(24^{+11}_{-3})$ MeV. Actually, these three dispersive values together with the one from the ``analytic K-matrix'' approach in \cite{Mennessier:2010xg}, $(981\pm43)-i(18\pm11)$ MeV, are the only new additions to the $f_0(980)$ in the 2012 RPP .

According to the RPP, the $K_0^*(800)$ or $\kappa$, still ``needs confirmation''. However, as with the $\sigma$, either within sounded models or rigorous Roy-like dispersive analyses \cite{DescotesGenon:2006uk}, it appears as a wide pole (not a Breit Wigner) around
650 to 770 MeV, with a 550 MeV width or larger. Moreover, as with the $\sigma$, a consistent pole is seen in heavy meson decays. Hence, to my view, it should be  treated on a similar footing as the $f_0(500)$ and considered as another ``well established'' light meson.

Unfortunately, the additions the 2012 RPP come from Breit-Wigner parameters of two studies of $J/\Psi$ decays at BES2 \cite{Ablikim:2010kd}. Fortunately this Collaboration \cite{Ablikim:2010kd} also provides a $t-matrix$ pole position
$764\pm^{+71}_{-54}-i(306\pm149^{+143}_{-85})$ MeV. This is quite consistent with the value from rigorous dispersive approaches $(658\pm13)-i(278.5\pm12)$ MeV \cite{DescotesGenon:2006uk}. The 2012 RPP keeps the previous estimates: $685\pm29$ MeV for the mass
 and $547\pm24$ for the width.

\vspace*{-.2cm}
\section*{Acknowledgments}
\vspace*{-.2cm}
Let me thank the organizers for creating such a nice opportunity to discuss Physics, and the group conveners for their invitation to present this review.
This work is supported in part
 by the Spanish Research contract , FPA2011-27853-C02-02 and  the EU FP7 HadronPhysics3 project.


\vspace*{-.3cm}
\begin{thebibliography}{99}
\vspace*{-.2cm}

\bibitem{PDG} J. Beringer et al. (Particle Data Group), Phys. Rev. D86, 010001 (2012). 

\vspace*{-.2cm}
\bibitem{Johnson:1955zz}
  M.~H.~Johnson and E.~Teller,
  Phys.\ Rev.\  {\bf 98}, 783 (1955).

\vspace*{-.2cm}
\bibitem{GellMann:1960np} 
  M.~Gell-Mann and M.Levy,
  Nuovo Cim.\  {\bf 16}, 705 (1960).

\vspace*{-.2cm}
\bibitem{Pr73}Protopopescu, S. D., et al., {\sl Phys Rev.} {\bf D7}, 1279, (1973).
Grayer, G., \emph{et al.},  {\sl Nucl. Phys.}  {\bf
B75}, 189, (1974).
Losty, M.~J., \emph{et al}.  {\sl Nucl. Phys.}, {\bf B69}, 185 (1974). Hyams, B.,\emph{ et al.}, {\sl Nucl. Phys.} {\bf B100}, 205, (1975).
  P.~Estabrooks and A.~D.~Martin,
  Nucl.\ Phys.\ B {\bf 79}, 301 (1974).
  R.~Kaminski, L.~Lesniak and K.~Rybicki,
  Z.\ Phys.\ C {\bf 74}, 79 (1997).

\vspace*{-.2cm}
\bibitem{Rosselet:1976pu}
  L.~Rosselet, {\it et al.},
  Phys.\ Rev.\  {\bf D15}, 574 (1977).
  S.~Pislak {\it et al.},  
  Phys.\ Rev.\ Lett.\  {\bf 87} (2001) 221801.

\vspace*{-.2cm}
\bibitem{Batley:2010zza}
  J.~R.~Batley {\it et al.} [ NA48-2 Collaboration ],
  Eur.\ Phys.\ J.\  {\bf C70}, 635-657 (2010).

\vspace*{-.2cm}
\bibitem{Amsler:1995bf} 
  C.~Amsler {\it et al.}  
  Phys.\ Lett.\ B {\bf 355}, 425 (1995) and
  Phys.\ Lett.\ B {\bf 342}, 433 (1995).


\vspace*{-.2cm}
\bibitem{Roy:1971tc} 
  S.~M.~Roy,
  Phys.\ Lett.\ B {\bf 36}, 353 (1971).


\vspace*{-.2cm}
\bibitem{CGL} 
  G.~Colangelo, J.~Gasser and H.~Leutwyler,
  Nucl.\ Phys.\ B {\bf 603}, 125 (2001).

\vspace*{-.2cm}
\bibitem{ACGL} 
  B.~Ananthanarayan, G.~Colangelo, J.~Gasser and H.~Leutwyler,
  Phys.\ Rept.\  {\bf 353}, 207 (2001).

\vspace*{-.2cm}
\bibitem{Kaminski:2002pe} 
  R.~Kaminski, L.~Lesniak and B.~Loiseau,
  Phys.\ Lett.\ B {\bf 551}, 241 (2003).


\vspace*{-.2cm}
\bibitem{Caprini:2005zr} 
  I.~Caprini, G.~Colangelo and H.~Leutwyler,
  Phys.\ Rev.\ Lett.\  {\bf 96}, 132001 (2006).


\vspace*{-.2cm}
\bibitem{GKPY11} 
  R.~Garcia-Martin,\emph{ et al.}, 
  Phys.\ Rev.\ D {\bf 83}, 074004 (2011).

\vspace*{-.2cm}
\bibitem{GarciaMartin:2011jx} 
  R.~Garcia-Martin, \emph{et al.}, 
  Phys.\ Rev.\ Lett.\  {\bf 107}, 072001 (2011).


\vspace*{-.2cm}
\bibitem{Moussallam:2011zg} 
  B.~Moussallam,
  Eur.\ Phys.\ J.\ C {\bf 71}, 1814 (2011).

\vspace*{-.2cm}
\bibitem{Pislak:2001bf}
  S.~Pislak {\it et al.}  [BNL-E865 Collaboration],
  Phys.\ Rev.\ Lett.\  {\bf 87} (2001) 221801.

\vspace*{-.2cm}
\bibitem{Mennessier:2010xg} 
  G.~Mennessier, S.~Narison and X.~G.~Wang,
  Phys.\ Lett.\ B {\bf 688}, 59 (2010).



\vspace*{-.2cm}
\bibitem{others}
  A.~Dobado and J.~R.~Pelaez,
  Phys.\ Rev.\ D {\bf 56}, 3057 (1997).
  J.~A.~Oller and E.~Oset,
  Phys.\ Rev.\ D {\bf 60} (1999) 074023.
  J.~A.~Oller, E.~Oset and J.~R.~Pelaez,
  Phys.\ Rev.\ D {\bf 59} (1999) 074001;
   [Erratum-ibid.\ D {\bf 60} (1999) 099906];
   [Erratum-ibid.\ D {\bf 75} (2007) 099903].
  J.~R.~Pelaez,
  Mod.\ Phys.\ Lett.\ A {\bf 19}, 2879 (2004)
  Z.~Y.~Zhou \emph{et al.}, 
  JHEP {\bf 0502}, 043 (2005).
  R.~Garcia-Martin and J.~R.~Pelaez, F.~J.~Yndurain, 
  Phys.\ Rev.\ D {\bf 76}, 074034 (2007)


\vspace*{-.2cm}
\bibitem{DescotesGenon:2006uk} 
  S.~Descotes-Genon and B.~Moussallam,
  Eur.\ Phys.\ J.\ C {\bf 48}, 553 (2006).



\vspace*{-.2cm}
\bibitem{Ablikim:2010kd} 
  M.~Ablikim, 
{\it et al.},
  Phys.\ Lett.\ B {\bf 693}, 88 (2010) and
  Phys.\ Lett.\ B {\bf 698}, 183 (2011).




\end{thebibliography}
\end{document}